%% file: main.tex
\documentclass[5p]{elsarticle}

\usepackage[utf8]{inputenc}
\usepackage{graphicx}
\usepackage{xcolor}
\definecolor{refC}{rgb}{0,0,0.75}
\usepackage[unicode,colorlinks,linkcolor=refC]{hyperref}

\usepackage{import}
\usepackage{multirow}
\usepackage{amsmath}

\usepackage{booktabs}

\usepackage{listings}
\usepackage{float}
\usepackage{stfloats}
\usepackage{placeins}

\newcommand*{\fullref}[1]{\hyperref[{#1}]{\textbf{\autoref*{#1}: \nameref*{#1}}}}
\newcommand*{\shortref}[1]{\hyperref[{#1}]{\ref*{#1} \nameref*{#1}}}

\journal{Computer Communications}

\begin{document}

\begin{frontmatter}

\title{Testing Implementation of FAMTAR: Adaptive Multipath Routing}

\author{Piotr Jurkiewicz\corref{mycorrespondingauthor}}
\cortext[mycorrespondingauthor]{Corresponding author}
\ead{piotr.jurkiewicz@agh.edu.pl}

\author{Robert Wójcik}
\author{Jerzy Domżał}
\author{Andrzej Kamisiński}

\address{Department of Telecommunications, AGH University of Science and Technology, Kraków, Poland}

\begin{abstract}
Flow-Aware Multi-Topology Adaptive Routing (FAMTAR) is a new approach to multipath and adaptive routing in IP networks which enables automatic use of alternative paths when the primary one becomes congested. It provides more efficient network resource utilization and higher quality of transmission compared to standard IP routing. However, thus far it has only been evaluated through simulations. In this paper we share our experiences from building a~real-time FAMTAR router and present results of its tests in a physical network. The results are in line with those obtained previously through simulations and they open the way to implementation of a production grade FAMTAR router.
\end{abstract}


\end{frontmatter}

% -----------------------------------------------------------------------------------------------------------------------------

\section{Introduction}
{FAMTAR} (Flow-Aware Multi-Topology Adaptive Routing) is a new approach to routing in IP networks, which provides multipath and adaptive capabilities. It is based on flows, notion currently incorporated in many architectures, including Software-Defined Networking (SDN). Unlike many SDN solutions, FAMTAR is a \textbf{fully distributed} mechanism and does not depend on a central entity. This is a key characteristic, as it significantly improves scalability and resilience.

Our aim was to test the efficiency of FAMTAR in real network conditions and validate the results obtained earlier through network simulations. In order to do that, we needed to build a FAMTAR router. In this paper we present our experiences in building a FAMTAR router. We also show the results of performance evaluation of networks built with FAMTAR routers. In particular:

\begin{itemize}
    \item We explain how to build a FAMTAR router using the Click Modular Router \cite{click-acm} framework, provide its configuration and describe key components.
    \item We identify and solve several problems, which were not noticed during simulations:
          \begin{itemize}
            \item The problem that a routing protocol may not be able to update routing tables on time after a link failure and flow entry deletion, which can result in subsequent packets being routed to a failed link.
            \item The problem with overwhelming generation of ICMP Redirect messages.
          \end{itemize}
    \item We show that, with FAMTAR, the amount of successfully transmitted data increases almost linearly with the number of available parallel paths (also, fewer packets are dropped in the network).
    \item We show that the average transmission delay in a FAMTAR-enabled network decreases with the increasing number of active parallel paths and is lower than in a standard IP network.
    \item These results are valid both for artificially generated traffic and real traffic from a BitTorrent network.
    \item We show that FAMTAR can effectively protect QoS parameters of VoIP flows and eliminate network congestions by redirecting excessive flows to alternative paths.
    \item We confirm that the TTL-based loop resolution mechanism works as expected, i.e., it resolves permanent loops which may appear in the network due to failures.
    \item Finally, we make the implementation publicly available as an open source software: \url{https://github.com/piotrjurkiewicz/famtar}
\end{itemize}

%%%%%%%%%%%%%%%%%%%%%%%%%%%%%%%%%%%%%%%%%%%%%%%%%%%%%%%%%%%%%%%%%%%%%%%%%%%%%%%%%%%%%%%%%%%%%%%%%%%%%%%%%%%%%%%%%%%%%%%%%%%%%%%%%%%%%%%%%%%%%%%%%%%%%%%%%%%%%%%%%
%%%%%%%%%%%%%%%%%%%%%%%%%%%%%%%%%%%%%%%%%%%%%%%%%%%%%%%%%%%%%%%%%%%%%%%%%%%%%%%%%%%%%%%%%%%%%%%%%%%%%%%%%%%%%%%%%%%%%%%%%%%%%%%%%%%%%%%%%%%%%%%%%%%%%%%%%%%%%%%%%
%%%%%%%%%%%%%%%%%%%%%%%%%%%%%%%%%%%%%%%%%%%%%%%%%%%%%%%%%%%%%%%%%%%%%%%%%%%%%%%%%%%%%%%%%%%%%%%%%%%%%%%%%%%%%%%%%%%%%%%%%%%%%%%%%%%%%%%%%%%%%%%%%%%%%%%%%%%%%%%%%
\section{Related work} \label{sec:background}

\subsection{Multipath routing}

There are numerous approaches to providing multipath transmissions in IP networks. Multipath solutions can also operate at different layers. The most visible solutions were surveyed in \cite{NCN_multipath}.

In the transport layer, Multipath TCP (MPTCP) is the most popular multipath solution. It is implemented in the Linux kernel, FreeBSD and Apple's iOS. It is crucial, because transport layer multipath requires support from both endpoints. MPTCP works by multiplexing a single TCP connection over multiple IP interfaces. Assuming that these interfaces provide disjoint paths to the other endpoint, MPTCP can increase throughput to the sum of available paths throughput. It also improves resilience, as any of the subflows can be disrupted, without loosing the main TCP connection (seamless takeover).

As mentioned, MPTCP requires existence of multiple interfaces on both endpoints. It is most beneficial, when these interfaces are connected to different upstream providers (multihoming). This is an uncommon case, so usage of MPTCP is rather limited. It is used in mobile scenarios for applications requiring low latency (e.g. voice assistant), when both cellular and WiFi connections are available. Another use case is bulk data transfer between datacenters. QUIC protocol, which is envisioned as future HTTP transport protocol and already makes up significant amount of Internet traffic, was also designed to support multipath. Details of this feature were postponed to subsequent protocol versions (as of February 2019), but it will probably face the same limitations as MPTCP.

Network layer multipath solutions do not require any support from endpoints. They require, however, additional features on routers, usage of specific protocols in the network or manual configuration from a network operator.

Equal cost multi-path (ECMP) is a standard multipath load balancing technique. It is enabled by default in many devices and can be used with majority of routing protocols. It utilizes the fact that routing protocol may find multiple shortest paths (all with the same cost) to a given destination. Traffic is then distributed equally amongst multiple shortest path next hops in each router. While existence of multiple shortest paths is common in highly-symmetric datacenter networks, it is rare in wide-area networks. For example in the \emph{nobel\_eu} topology from SNDLib (\url{http://sndlib.zib.de}), average number of disjoint shortest paths is $1.20$, whereas average number of disjoint paths (max-flow/min-cut) equals to $2.61$. Lack of multiple shortest paths significantly reduces benefits from ECMP in wide-area networks.

Unequal cost multi-path (UCMP) assumes usage of additional paths with cost higher than the shortest one's cost. This is not trivial, as it may lead to routing loops. Loop-free UCMP requires specific metrics and constrains in routing protocol. The only protocol currently in usage supporting UCMP is EIGRP. Its conservative feasibility conditions (the DUAL algorithm), however, significantly limit the number of additional paths possible to use \cite{dual}. In order to use all available disjoint paths and achieve maximum flow between selected nodes, other techniques have to be used.

There are also many UCMP algorithms, which were proposed in academia, but have not been implemented in any routing protocol. This includes LFI-type algorithms: MPDA and MDVA, both proposed by the DUAL algorithm inventor \cite{vutukury2001multipath}, and OSPF extensions: OMP \cite{draft-ietf-ospf-omp-02} and AMP \cite{gojmerac2007adaptive}, which are limited by a similar conservative feasibility conditions as DUAL.

\subsection{Adaptive routing}

Per-packet routing imposes significant limitations not only on multipath capabilities (due to routing loop prevention constrains, only a subset of existing alternative paths in the network can be used), but also on adaptivity. Adaptive (load-sensitive) routing is impossible in the per-packet approach, as the dynamic alteration of link costs leads to instability, which ultimately deteriorates network performance. This has been shown by early ARPANET attempts.

The original ARPANET routing algorithm used the current link delay as a metric \cite{mcquillan1978review}. It consisted of propagation and transmission times, which were constant, and variable queuing delay, which depended exponentially on the link usage. Its successor (\emph{The New ARPANET Routing Algorithm}), which was a link-state routing protocol, used a 10-seconds average link delay as a metric \cite{Mcquillan80thenew}. This metric was called \emph{Delay Shortest Path First (D-SPF)}. The new algorithm allowed to avoid permanent routing loops and was more stable. However, under high load, it still resulted in route oscillations. In July 1987 the metric was normalized in order to depend linearly on the link usage (\emph{Hop-Normalized Shortest Path First (HN-SPF)}). This resulted in a more smooth handover of traffic from overloaded path, but still did not eliminate the route flapping \cite{kz89} \cite{zinky1989performance}.

These early ARPANET adaptive routing stability problems were analysed in \cite{Bertsekas82}. It was proven in \cite{wang1992analysis} that in some cases adaptive routing can degrade the overall network performance even below the non-adaptive routing. Since that, approaches based on adaptive routing protocol metric changes were abandoned. However, some remnants of them are still present in the currently used routing protocols. For example, EIGRP can make use of link load in metric calculation, but this option is turned off by default \cite{igrp-stability}.

Another approach was explored in an OSPF extension called OMP \cite{draft-ietf-ospf-omp-02}. The authors proposed to change load-balancing weights instead of routing protocol metrics in response to overload events. They suggested to disseminate load information along with LSA messages. AMP \cite{gojmerac2007adaptive} was a similar proposal. The difference was that the information was exchanged directly between neighbor routers using so-called \emph{backpressure messages}. Both these extensions have not been implemented yet and their stability under a realistic traffic have not been evaluated.

In practice, multipath and adaptive routing is currently achieved with the help of Multiprotocol Label Switching (MPLS) \cite{rfc3031}. MPLS allows operators to manually create paths (tunnels) and assign certain flows/transmissions to those paths. This approach is currently widely used, and it works. However, there are several drawbacks. Firstly, the operations are not automatic and require human intervention. Secondly, in large and complex networks the existence of multiple paths and many criteria, conditions, parameters, etc., creates havoc. Currently, many major operators have their MPLS nodes configured with such complexity that they are almost afraid to change anything, for fear of impairing the network's operation.

There are proposals of automated systems, which purpose is to modify MPLS tunnels in a response to the changing network load. Usually these systems assume usage of a centralized controller, which makes them incomparable with FAMTAR. However, distributed algorithms, like MATE \cite{mate} or TeXCP \cite{texcp}, were also proposed. They collect network load information using probe packets, but no implementations are available.

Non-MPLS based mechanisms were also proposed, for example REPLEX \cite{replex}. However, it is complicated and introduces additional signaling protocol to the network. Another similar solution is proposed in \cite{kvalbein2009multipath}, but its drawback is that it makes routing decisions basing only on local interfaces load, instead of the whole network state. A short survey of adaptive routing approaches is presented in \cite{skrypnyuk2006load}.

\subsection{Flow-based routing}

FAMTAR was developed to answer the aforementioned problems of per-packet routing by making the use of flow-based approach. It can overcome multipath and adaptivity limitations of per-packet routing by maintaining separate per-flow forwarding entries. As a result, flows between the same endpoints can follow any number of alternative paths without the risk of loops. Furthermore, paths for subsequent flows can be chosen with the current or a predicted network load in mind, effectively resulting not only in the multipath routing, but also in the adaptive routing. Such an approach should also ensure a better stability, as only new flows can be redirected to new paths during a congestion, which can reduce route flapping.

Caspian Networks and, later, Anagran tried to provide flow-based treatment. In {\cite{flow_routing}}, it is shown that keeping flow state information is feasible. Moreover, Anagran created FR-1000, a router which provided flow-based treatment and could be used for high speed links. Anagran stores packet forwarding information inside flow tables, but unlike FAMTAR, this is not modified according to network congestions. FAMTAR uses similar flow routing information to that used in Anagran, and combines it with routing adaptability to the current network congestion statuses.

A relatively new proposal for flow management was presented in \cite{rubin2010max}. In this solution, flows are classified and transmitted using multiple paths. A central manager decides which paths should be used for each flow. This proposal looks promising, however, it is complex and difficult to implement. Moreover, the existence of the central manager may result in scalability and security problems in large networks.

Central management is also the tendency of currently sound Software-Defined Networks (SDN). There are hundreds of papers proposing SDN flow-based traffic engineering and routing systems. Some of these systems are fairly sophisticated, utilizing recent AI and machine learning advances \cite{aite}. All of them are (at least logically) centralized. Therefore, we purposefully do not review and compare them to FAMTAR, which is a fully distributed solution and thus belongs to a completely another class of algorithms.

\subsection{FAMTAR origins and possible extensions}

The development of FAMTAR was possible thanks to the advances in flow-aware traffic handling, such as Flow-Aware Networking (FAN) \cite{FAN_first}. There were attempts to increase the efficiency of routing with the use of FAN, as presented, for example, in \cite{FAN_routing}. The technique of trunk reservation borrowed from the telephone network is proposed in this paper for route selection. It is assumed that a path for a flow is chosen based on the bandwidth it requires. Also, a simple intelligent routing for FAN is presented in \cite{intelligent_FAN}, where it is assumed that only non-congested links are considered when forwarding packets. However, it is not specified how to inform all routers about the congested links. FAMTAR provides a solution to that problem. Moreover, FAMTAR is a general idea, not specific to FAN networks.

The basic implementation of FAMTAR can be extended with other existing solutions, like admission \cite{FAMTAR_AC} and congestion control approaches or source-based routing. Here we present some examples of mechanisms and algorithms which can be considered for future research.

In the paper \cite{zhang2019optimized} a delay response BBR (Bottleneck Bandwidth and Round-trip time) algorithm designated for real time video transmission is proposed and analyzed. The main assumption of the proposed algorithm is to reduce sending rate when the link delay exceeds predefined threshold. The goal is to allow routers to maximize the usage of bandwidth by control of buffer occupation. The sending rates of flows are being observed and actively reduced when delay exceeds a predefined threshold. This allows a router to drain queued buffers and, as a consequence, to reduce packet delay and loss rate. The authors proved by simulation experiments that the proposed congestion control algorithm achieves lower frame delivery delay in comparison to benchmark algorithms. The algorithm can be implemented in any multipath architecture to enable maximum bandwidth utilization and to ensure quality of service at the assumed level for selected traffic. This proposal can also be implemented in FAMTAR, which originally has been proposed to serve traffic in a best effort regime.

An approach similar to the described above has been proposed in \cite{Cominetti2014}. The authors present a model which combines a Network Utility Maximization to control rate of flows taking into account end-to-end queuing delays. However, in this solution, a Markovian Traffic Equilibrium is used to decide on routing based on total expected delays. The authors explain that the proposed method uses a routing strategy which is based on a decentralized stochastic version of Wardrop's model. Provided theoretical analysis confirm that it is possible to establish decentralized multipath routing algorithm which control congestions. The presented solution has been described only theoretically. The authors did not show simulation or tests of hardware implementation results. The proposed solution is more complex than FAMTAR. It needs additional signaling to distribute information about queuing delays in each outgoing link of all network nodes. One of the main advantages of FAMTAR is lack of any additional signalization in a network.

Another congestion control approach for multipath implementation has been proposed in \cite{Mahdian:2016:MMI:2984356.2984365}. The authors present a rate-based, Multipath-aware Information-centric networking Rate-based Congestion Control (MIRCC) approach. The proposed solution is inspired by the Rate Control Protocol (RCP). MIRCC ensures an acceptable convergence time with less overshoot and oscillation in comparison to the RCP and a multipath transmission along all the available paths, maintaining fairness among active competing flows regardless of number of paths that each flow has.

The MIRCC approach requires an implementation of several algorithms and protocols, e.g. algorithm in each forwarder calculating dual-class rates for each link, protocol mechanisms to communicate rates and path identifiers to consumers in data messages, an algorithm to determine interest sending rates for each class and to determine a sensible distribution of interests across available paths. As one can see, the proposed mechanism is complex and has been proposed for ICN (Information-Centric Networking). Thus the implementation of the described approach in FAMTAR is outside the scope of this paper.

Recently proposed by IETF, RFC 8354 \cite{rfc8354}, presents use cases for IPv6 Source Packet Routing in Networking (SPRING). SPRING is a networking architecture which leverages the source routing paradigm.  The main assumption is that an ingress node steers a packet by including a controlled set of instructions, called segments, in the SPRING header. The mentioned architecture can be used in e.g. small offices, access networks or data-center networks. Of course, it can also be implemented in core networks. The authors list that it can enable to e.g.:

\begin{itemize}
\item use selected high-bandwidth links for a specific type (high priority) of traffic and thus avoid the need for overdimensioning the links in the network,
\item setup separate path for delay-sensitive flows,
\item reach important servers through source-based optimal path,
\item use of disjoint paths.
\end{itemize}

Source-routing approaches are promising solutions for multipath routing of flows. They, however, need additional signaling comparing with FAMTAR. The architecture like SPRING can be implemented in network with FAMTAR. Both architectures can complement each other. However, at this moment, we focus on basic FAMTAR implementation to show its effectiveness and simplicity.

\section{Flow-Aware Multi-Topology Adaptive Routing} \label{sec:FAMTAR}
FAMTAR was introduced in \cite{FAMTAR_LETTERS}. It is a multipath adaptive routing mechanism that works based on flows \cite{ACM_flow-oriented}. FAMTAR is placed above the IGP (it does not interfere with the routing protocol operation) and can work with every protocol. A routing protocol is responsible for finding the best path between two endpoints up to its capabilities. In an uncongested network, all transmissions between those endpoints use this path. When a path becomes congested, all new flows are pushed to an alternative path, while flows which are already active remain on their primary path. Therefore, FAMTAR uses the best path provided by the routing protocol, and automatically triggers finding new paths in case of congestion.

To achieve that, every FAMTAR router maintains a Flow Forwarding Table (FFT) alongside the classic routing table. FFT is an associative array in which the keys are flow identifying fields. For each flow, the corresponding FFT entry indicates the interface to which packets of this flow are forwarded. This information is taken from the current routing table when the flow is added to the FFT, i.e., when its first packet appears. Entries in the FFT are static and do not change alongside routing table changes. FFT is used to realize most packet routing tasks, as for flows that are present in the FFT, the routing table is not consulted.

\begin{figure}[!h]
\centering
\includegraphics[width=0.8\columnwidth]{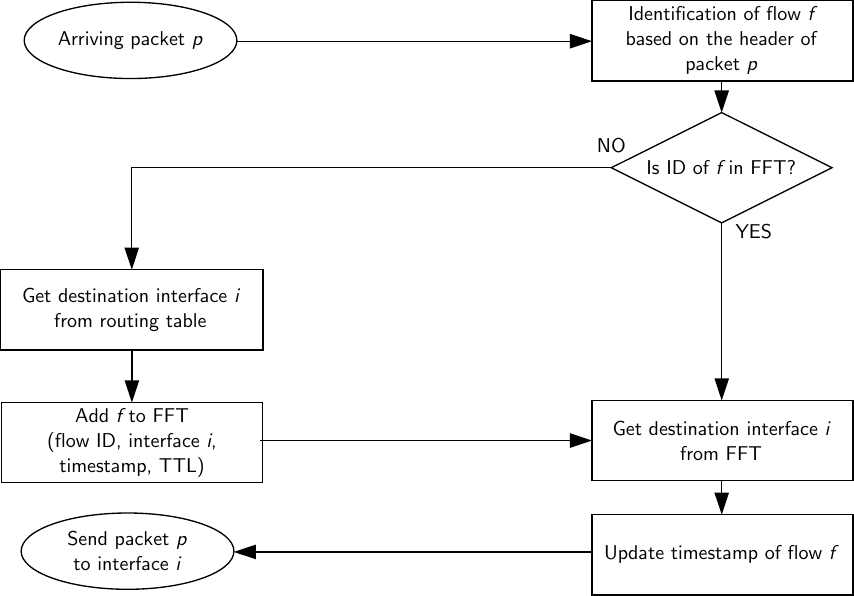}
\caption{Block diagram of FAMTAR algorithm.}
\label{algo_basic}
\end{figure}

When a state close to congestion is noticed on one of the links, the corresponding router sets the cost of this link to a predefined high value. From this moment, this link is perceived as congested. The new cost appears as a change in the routing protocol, which disseminates this information as a standard topology change message. Upon receiving this information, routing protocols compute new paths which are likely to avoid congested links. The newly computed paths cause the related changes in the routing tables. However, these changes affect only new flows. The old flows, which were active before that event, are still routed on their existing paths stored in the FFT entries. Although the congested link still forwards all the flows which were active before the congestion was noticed, new transmissions do not appear. After a while, when the congestion on this link stops, the original cost of the link is restored. Note that FAMTAR requires a router to detect congestion on one of its links. The method to determine the congestion is not specified, although any congestion indicator can be used (e.g., link load, queue occupancy, packet queuing delay, and so on). The block diagram of FAMTAR algorithm is presented in \autoref{algo_basic}.

The key idea of FAMTAR is keeping the system distributed by leveraging the IGP to perform route computations instead of communicating with a central controller. Moreover, unlike traditional best-effort approaches, which usually can utilize only multiple shortest paths (ECMP), FAMTAR can use all available paths between transmitting nodes. The \autoref{tab:topocomp} presents statistics of three real WAN topologies from the SNDLib database (\url{http://sndlib.zib.de}). The number of disjoint paths between nodes is 2-3 times greater than the number of disjoint shortest paths. Usage of these additional paths is impossible in ECMP. Unequal cost multipath algorithms (UCMP), such as EIGRP, allow to use some of these additional paths, but their set is still very limited by a conservative feasibility condition. FAMTAR's flow-based approach makes it possible to overcome these constrains and utilize all existing paths. This increases the efficiency of network resource utilization and reduces the need for over-provisioning, since it is possible to use all already available resources.

Another key factor, which distinguishes FAMTAR from the other multipath approaches, is the usage of alternative (and potentially suboptimal) paths only when necessary and only to new transmissions. A path with a higher cost is used only when paths with lowest costs are congested, whereas EIGRP uses all paths regardless of their load, which means that some packets are routed on worse paths even when there is no congestion on the best path. In FAMTAR, upon congestion, only new flows use alternative paths, whereas the old ones remain on their primary paths.

Flow-based approach also improves stability and reduces route flapping, as only the new flows are forwarded to new paths. Traditional adaptive routing approaches (such as ARPANET routing algorithms) had stability issues, because when the original path was becoming congested, the whole traffic was switched to the alternative path, which resulted in oscillations between these two paths. Simulation results presented in \cite{FAMTAR_CQR} show that FAMTAR is able to significantly increase the amount of traffic sent in a network, while reducing packet delays.

\begin{table}[!h]
\centering
\scriptsize
\caption{Statistics of selected SNDLib real WAN topologies} \label{tab:topocomp}
\vspace*{6pt}
\begin{tabular}{@{}lrrr@{}}
\toprule
Topology name                          & polska	& nobel\_eu	& germany50 \\
\midrule
Number of nodes                        & 12	& 28	& 50 \\
Number of edges                        & 18	& 41	& 88 \\
Graph density                          & 0.27	& 0.11	& 0.07 \\
Avg. vertex degree                     & 3.00	& 2.93	& 3.52 \\
Avg. num. of disjoint shortest paths   & 1.27	& 1.20	& 1.23 \\
Avg. num. of all disjoint paths        & 2.67	& 2.61	& 3.19 \\
\bottomrule
\end{tabular}
\end{table}

%%%%%%%%%%%%%%%%%%%%%%%%%%%%%%%%%%%%%%%%%%%%%%%%%%%%%%%%%%%%%%%%%%%%%%%%%%%%%%%%%%%%%%%%%%%%%%%%%%%%%%%%%%%%%%%%%%%%%%%%%%%%%%%%%%%%%%%%%%%%%%%%%%%%%%%%%%%%%%%%%
%%%%%%%%%%%%%%%%%%%%%%%%%%%%%%%%%%%%%%%%%%%%%%%%%%%%%%%%%%%%%%%%%%%%%%%%%%%%%%%%%%%%%%%%%%%%%%%%%%%%%%%%%%%%%%%%%%%%%%%%%%%%%%%%%%%%%%%%%%%%%%%%%%%%%%%%%%%%%%%%%
%%%%%%%%%%%%%%%%%%%%%%%%%%%%%%%%%%%%%%%%%%%%%%%%%%%%%%%%%%%%%%%%%%%%%%%%%%%%%%%%%%%%%%%%%%%%%%%%%%%%%%%%%%%%%%%%%%%%%%%%%%%%%%%%%%%%%%%%%%%%%%%%%%%%%%%%%%%%%%%%%
\section{Implementation environment} \label{sec:environment}
To implement FAMTAR, we had to find a router environment which would allow modifications in packet processing. This environment had to be expanded by introducing FFT and all actions related to using this table. Therefore, the implementation environment for an experimental FAMTAR router must allow easy modifications and debugging of the packet processing path. This requirement led us to choose the Click Modular Router suite (Click) as a data plane provider.

Click is a suite for building flexible software packet processors, designed with research and experimental applications in mind. It was developed by Kohler \cite{click-thesis} in his Ph.D. dissertation. Click is widely used for building experimental software routers and switches. Its advantages include considerable flexibility, clear and scalable architecture, ease of adding new features, and high performance.

Click achieves flexibility due to its modular and object-oriented architecture. Routers are assembled from fine-grained packet processing modules called elements, which are C++ classes. Each individual element performs a simple operation on a packet, like queuing or decrementing a packet's time to live (TTL) field. Each element has input and output ports, which serve as the endpoints of connections between them. A user builds a complete router configuration by connecting individual elements into a directed graph. During router operation, packets are processed sequentially by individual elements.

Click can run as a user-level application or as a Linux kernel module. In the kernel mode, the Click module replaces the operating system (OS) networking stack, and packet processing is done only by Click. In the user-level mode, Click uses the system to receive packets.

Choosing Click as the data plane platform determined another router components including the routing daemon. Since Click does not implement dynamic routing protocols, its routing table must be populated by an external routing daemon. The only daemon that cooperates with Click is XORP (eXtensible Open Router Platform) \cite{XORP}. XORP is an open source IP routing software suite originally designed at the International Computer Science Institute in Berkeley, California. It supports various routing protocols, including OSPF, BGP and RIP. Click combined with XORP provide all the required functions that each router supports, and this was the point in which our FAMTAR implementation started.

%%%%%%%%%%%%%%%%%%%%%%%%%%%%%%%%%%%%%%%%%%%%%%%%%%%%%%%%%%%%%%%%%%%%%%%%%%%%%%%%%%%%%%%%%%%%%%%%%%%%%%%%%%%%%%%%%%%%%%%%%%%%%%%%%%%%%%%%%%%%%%%%%%%%%%%%%%%%%%%%%
%%%%%%%%%%%%%%%%%%%%%%%%%%%%%%%%%%%%%%%%%%%%%%%%%%%%%%%%%%%%%%%%%%%%%%%%%%%%%%%%%%%%%%%%%%%%%%%%%%%%%%%%%%%%%%%%%%%%%%%%%%%%%%%%%%%%%%%%%%%%%%%%%%%%%%%%%%%%%%%%%
%%%%%%%%%%%%%%%%%%%%%%%%%%%%%%%%%%%%%%%%%%%%%%%%%%%%%%%%%%%%%%%%%%%%%%%%%%%%%%%%%%%%%%%%%%%%%%%%%%%%%%%%%%%%%%%%%%%%%%%%%%%%%%%%%%%%%%%%%%%%%%%%%%%%%%%%%%%%%%%%%
\section{Router components} \label{sec:components}

\begin{figure}[!h]
\centering
\includegraphics[width=0.8\columnwidth]{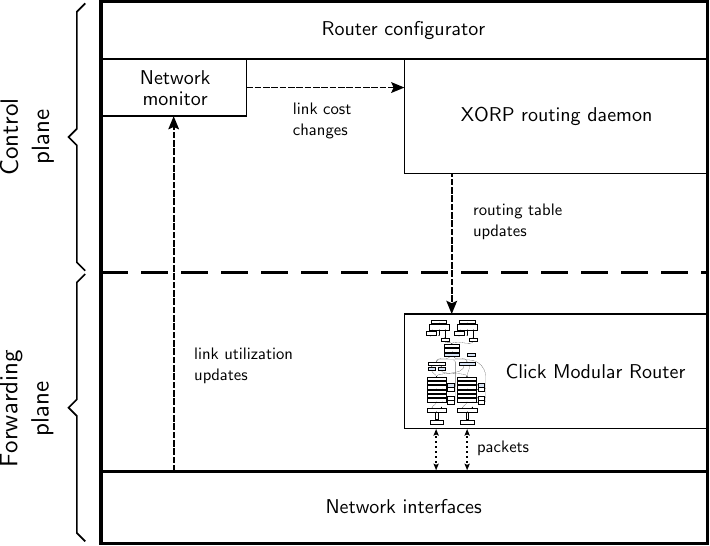}
\caption{FAMTAR router architecture on the Linux OS.}
\label{famtar_router}
\end{figure}

\autoref{famtar_router} presents the FAMTAR prototype router build on top of Linux Debian. The system runs Click and XORP with our extensions as well as auxiliary scripts written in Python. Click is responsible for forwarding packets and realizing FFT functions. Its routing table is updated by XORP which runs an OSPF protocol. Any link cost change in a network triggers XORP functions which results in routing table updates in Click.

The network monitor collects load information from physical interfaces. The monitor changes link costs in XORP for links whose traffic load exceeds the congestion threshold. The analysis of link loads is performed only a few times per second. Therefore, the implementation of this mechanism does not need to be extremely effective, so we implemented it in an external script.

The FAMTAR router configurator, which is a graphical front-end for configuration of the XORP daemon and the monitor, allows configuring the IP addresses of interfaces, congestion thresholds, and other router parameters.

\begin{figure}[h!]
\centering
\includegraphics[width=0.94\columnwidth]{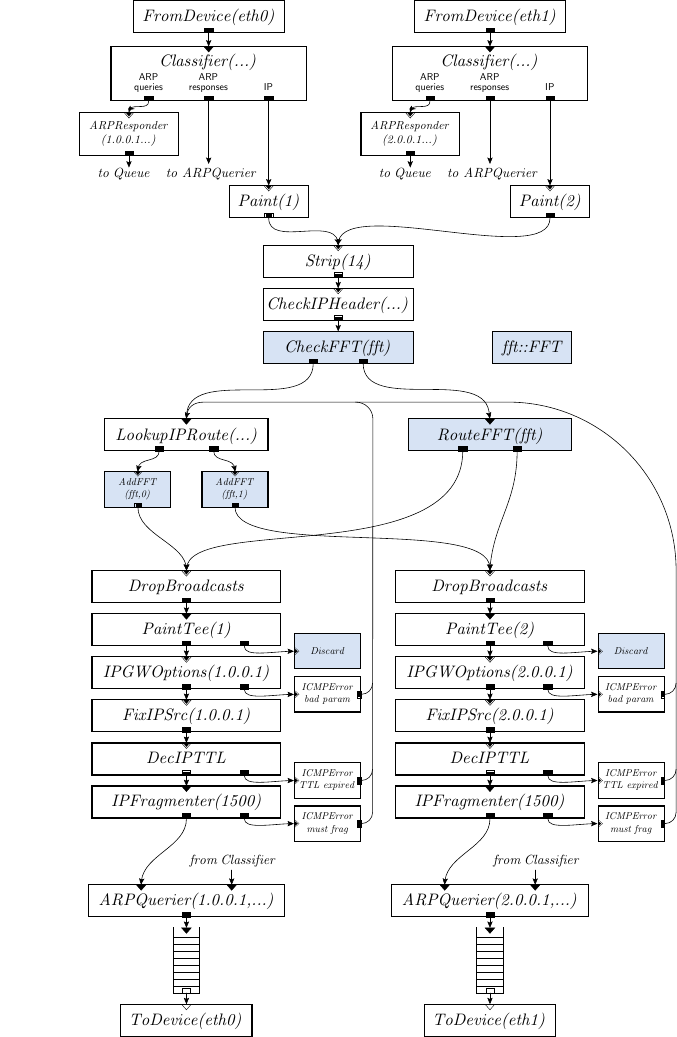}
\caption{Block diagram of FAMTAR implementation in Click Modular Router. Blocks marked in blue are related to FAMTAR and were added to Click.}
\label{click_router}
\end{figure}

The most important element, however, is the extension of Click. To implement FAMTAR in Click, it was necessary to add a few additional elements to the standard IP router. The block diagram of the FAMTAR router in Click is presented in \autoref{click_router}. This graph is an extension of the standard IP router graph, presented in \cite{click-acm}. The blocks related to FAMTAR are marked in blue shading. New elements: \emph{CheckFFT}, \emph{AddFFT}, \emph{RouteFFT}, and auxiliary element \emph{FFT} are not part of the Click library. They are written by us and grouped in the Click package \texttt{famtar} \cite{github-famtar}. This package may be downloaded and imported to Click as a shared library without Click modification or recompilation.

The first FAMTAR-related element is \emph{CheckFFT}. It is placed before the \emph{LookupIPRoute} block, where routing is executed. The aim of the \emph{CheckFFT} element is to check whether the FFT contains an entry for the flow related to the incoming packet. If the FFT does not contain the flow entry (i.e., it is a new flow), the incoming packet is sent to the first outgoing port of the \emph{CheckFFT} element. Next, this packet is served in the \emph{LookupIPRoute} element and sent to the correct outgoing port based on the current routing table. After this operation, the new entry for the particular flow is added to the FFT in the \emph{AddFFT} element. The entry includes the incoming time of the packet, its TTL value, the identifier of the outgoing port and the IP address of next router on the path.

When the active flow entry exists, the timestamp in this flow is updated and the incoming packet is sent to the second outgoing port of the \emph{CheckFFT} element. Then, the packet is processed by the \emph{RouteFFT} element. This element operates similarly to \emph{LookupIPRoute} and is responsible for packet routing. However, the packet is directed based on the outgoing port number stored in the FFT entry, instead of the one from the current routing table.

Additional operation is executed in the \emph{AddFFT} element, when a failure occurs in the network. When a router detects lack of carrier on the outgoing link, the \emph{AddFFT} element blocks for a fixed period (we assumed 5 s) a possibility to add new flows for that particular link. This operation is necessary for the correct implementation of the loop-resolution mechanism proposed in \cite{FAMTAR_TTL}. This mechanism assumes that after a link failure, all flow entries for this link are deleted from FFT. However, it takes some time to calculate new paths by the routing protocol. In this period, new flows outgoing via the recently-failed interface cannot be accepted to FFT. This means that during this period, these flows are being routed using solely the current routing table.

Moreover, it was necessary to switch off the mechanism implemented in the standard IP router, which sends error ICMP Redirect messages to the source node when a packet is rerouted to the same interface it arrived from. This mechanism generates traffic which is not necessary in FAMTAR network, because possible loops are solved by the TTL-based mechanism. To switch off this mechanism in Click, we changed the \emph{ICMPError} element with \emph{Discard} element presented in \autoref{click_router}.

All described elements: \emph{CheckFFT}, \emph{AddFFT} and \emph{RouteFFT} operate on the same FFT, which is implemented in the auxiliary element known as \emph{FFT}. This element is not placed directly on the packet processing path. However, it offers the other elements some functions to modify the FFT. The structure of key for FFT is as follows:

\begin{lstlisting}[basicstyle=\ttfamily\footnotesize,label=lst:flowkey]
struct FlowKey
{
   uint32_t srcaddr;
   uint32_t dstaddr;
   uint16_t srcport;
   uint16_t dstport;
   uint8_t  ip_prot;
}
\end{lstlisting}

FFT stores time of last packet of a flow, routing information and TTL value of the first packet of the flow. The structure of values stored in the flow forwarding table is presented below:

\begin{lstlisting}[basicstyle=\ttfamily\footnotesize,label=lst:flowval]
struct FlowValue
{
   uint32_t ts;
   uint8_t port;
   IPAddress gateway;
   uint8_t ttl;
}
\end{lstlisting}

Each flow key contains $104$ bits ($13$ bytes) in which flow identification fields are stored: IPv4 source and destination addresses ($2 \cdot 32$ bits), source and destination transport layer port numbers ($2 \cdot 16$ bits), and transport layer protocol number ($8$ bits). Each flow entry contains the following information: the time of last packet ($32$ bits), forwarding interface ($8$ bits), IP address of the next hop router ($32$ bits), and the TTL value of the first packet of the flow ($8$ bits). The timestamp is used to determine whether the particular flow entry expired or not. If no packets arrive for a predefined period of time, the flow entry is deleted. The number of the outgoing interface and the IP address of the next router on the path are necessary to perform routing in the \emph{RouteFFT} element. The TTL value is required by the loop resolution mechanism. The total size of a flow entry is $80$ bits ($10$ bytes). The total amount of information required to store each flow is therefore $104$ bits for a key plus $80$ bits for data, which gives $184$ bits or $23$ bytes. This means that to process $1$ million simultaneous flows, a router needs $23$ MB of memory for FFT, which is acceptable.

The \texttt{HashTable<>} container from the standard Click library is used to implement this associative array as a hash chain array. This make it possible to ensure low computing complexity -- $O(1)$ -- for operations made on packets. When a flow finishes transmission, its identifier should be removed from FFT. While routers are not aware explicitly when flows finish transmission, we had to implement an additional mechanism -- garbage collection. The content of buckets is analyzed when new flow entries are added to them and these entries, whose timestamp is older than a predefined timeout, are removed from FFT

%%%%%%%%%%%%%%%%%%%%%%%%%%%%%%%%%%%%%%%%%%%%%%%%%%%%%%%%%%%%%%%%%%%%%%%%%%%%%%%%%%%%%%%%%%%%%%%%%%%%%%%%%%%%%%%%%%%%%%%%%%%%%%%%%%%%%%%%%%%%%%%%%%%%%%%%%%%%%%%%%
%%%%%%%%%%%%%%%%%%%%%%%%%%%%%%%%%%%%%%%%%%%%%%%%%%%%%%%%%%%%%%%%%%%%%%%%%%%%%%%%%%%%%%%%%%%%%%%%%%%%%%%%%%%%%%%%%%%%%%%%%%%%%%%%%%%%%%%%%%%%%%%%%%%%%%%%%%%%%%%%%
%%%%%%%%%%%%%%%%%%%%%%%%%%%%%%%%%%%%%%%%%%%%%%%%%%%%%%%%%%%%%%%%%%%%%%%%%%%%%%%%%%%%%%%%%%%%%%%%%%%%%%%%%%%%%%%%%%%%%%%%%%%%%%%%%%%%%%%%%%%%%%%%%%%%%%%%%%%%%%%%%
\section{Tests of the FAMTAR router} \label{sec:tests}
So far, the functionality of FAMTAR has been tested and the results have been presented only basing on simulation experiments conducted in the ns-3 simulator. They showed advantages of the FAMTAR routing over the standard IP routing. However, we have to note that the simulation analysis is usually provided with limitations. For example, not all factors from real networks can be taken into account in simulations.

The most valuable results can be obtained when a router is tested in a network where real users generate traffic. We can assume that a device which passes such tests works properly and is scalable. Unfortunately, it is difficult and risky to test a prototype in a production network. That is why we conducted our tests in a laboratory environment, using either real of artificially generated traffic. We prepared a small network, as presented in Figures~\ref{fig:scenario-1}, \ref{fig:scenario-2}, \ref{fig:scenario-3}, \ref{fig:scenario-4} and \ref{fig:scenario-5}. All links in our network were bidirectional with capacity equal to 10 Mbit/s. Only links between border routers and hosts had 100 Mbit/s capacity. We have chosen such a low capacities in order to make sure that results will not be influenced by insufficient computing power of PCs we were using as routers.

Congestion thresholds were set to $ Th_{min} = 0.7 $ and $ Th_{max} = 0.9 $, as previous simulation papers showed that these values provide the best compromise between stability and performance. Link usage counters were refreshed every 200 milliseconds and the exponential moving average with $ \alpha = 0.2 $ was used to smooth out the readings. Moreover, frequency of route table recalculations was also limited by OSPF hold timer (similar to BGP's route dampening), which in the case of XORP daemon used by us was equal to 1 second. The default routing protocol cost of all links was equal to 1 (with the exception of scenario II). When a congestion on a link was detected, its cost was being raised to 100.

Traffic was generated in H1, which was a PC computer. We used the D-ITG \cite{D_ITG} application to generate traffic and to collect statistical data. In scenario II and III, we did not use the D-ITG traffic generator. Instead, in scenario II, an UDP packet generator was used in order to achieve realistic flow length and size distributions. In scenario III, R1 was connected to the Internet and we observed traffic in the network in the case when the destination host was downloading a file from the Internet through the P2P protocol. We repeated each experiment five times to collect statistically credible results. In most scenarios we observed four cases, with one, two, three and four possible paths between edge routers R1 and R4. Each time, we compared the results obtained for routers with turned on FAMTAR mechanism with the results obtained for routers doing classic IP routing.

The following sections present the ideas and results from four scenarios that we analyzed.

% -----------------------------------------------------
% SCENARIO 1
%
% DO WERYFIKACJI:
%  * okres aktywnosci generatora ruchu: 250 sekund
%  * zbierane dane z zakresu od 30 do 230 sekundy
%  * rozmiar pakietu: 1000 B
%  * liczba przeplywow UDP: 500 (uruchamiane ~~w regularnych odstepach~~ wedlug rozkladu eksponencjalnego co 0.5 s w przedziale 0-250 s)
%  * rozmiar przeplywu z rozkladu Pareto: avg 1000000 B, shape 1.25
%  * stala szybkosc bitowa przeplywu: 100 kB/s
%
% ZWERYFIKOWANO -- PJ

\subsection*{Scenario I}
\begin{figure}[h!]
\centering
\includegraphics[width=1.0\columnwidth]{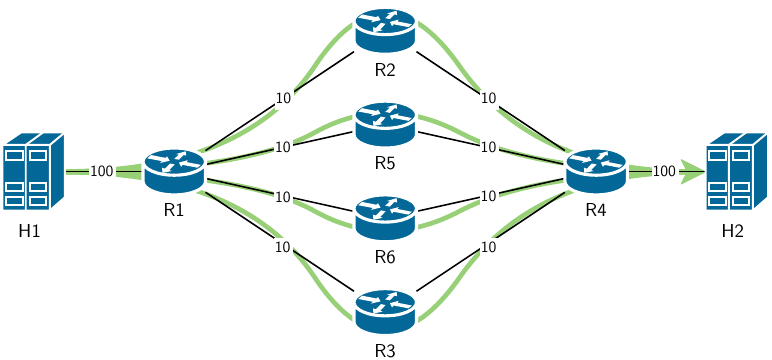}
\caption{Scenario I -- Testing topology.}
\label{fig:scenario-1}
\end{figure}

To verify the performance of FAMTAR with regard to its multipath capabilities, $500$ UDP flows were transmitted between nodes H1 and H2 (\autoref{fig:scenario-1}). Packet size was set to $1000$~B, whereas the flow size was selected according to the Pareto distribution (average: $1$~MB, shape: $1.25$). Each flow was transmitting data at a~constant rate of $100$~kB/s and the time between flow starts was given by the exponential distribution, with the average value of $0.5$ s. We collected the results between the 20th and 230th seconds of each experiment.

We conducted this experiment in networks with $1$, $2$, $3$, and $4$ parallel paths available. We assesed the performance basing on the following metrics:
\begin{itemize}
	\item number of bytes received at node H2,
	\item dropped packets ratio,
	\item average packet delay,
	\item minimum packet delay,
	\item maximum packet delay.
\end{itemize}

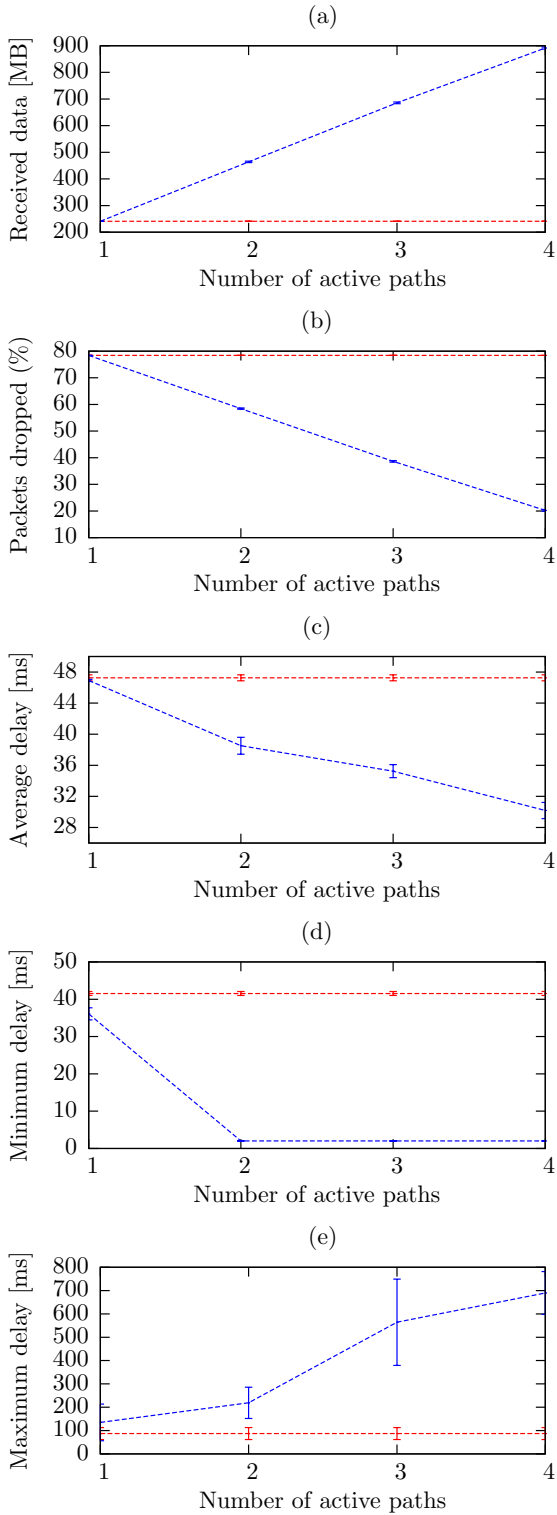
\begin{figure}[H] \center
	\scalebox{0.9}{\import{Plots/scenario-1/}{evaluation-scenario-1}}
	\caption{Scenario I -- Relations between the number of active paths and (a) the amount of received data, (b) the percentage of packets dropped, (c) the average transmission delay, (d) the minimum transmission delay, and (e) the maximum transmission delay in the standard IP and FAMTAR-enabled networks (red and blue lines, respectively).} \label{fig:evaluation-scenario-1}
\end{figure}

The results are presented in \autoref{fig:evaluation-scenario-1}. In the case when only one path is available, the results for the FAMTAR-enabled network are similar to the results for the standard IP network. This is in line with our expectations, since if no additional paths are available, FAMTAR cannot provide multipath transmissions and there is no gain.

In the case of a standard IP network, the results do not change with increasing number of additional paths. At the same time, in the case of the FAMTAR-enabled network, we note that the amount of successfully transmitted data increases linearly with the number of parallel paths. In addition, fewer packets are dropped in the network. This observation confirms one of the most significant advantages of FAMTAR, which is the ability to efficiently provide parallel multipath transmissions.

We also note that the average transmission delay in a~FAMTAR-enabled network decreases with the increasing number of active parallel paths and is lower than in the case of a standard IP network.
% At the same time, the maximum delay values increase with the number of parallel paths in FAMTAR, what means that significant delay fluctuations may occur.

\begin{figure}[h!]
\centering
\includegraphics[scale=0.9]{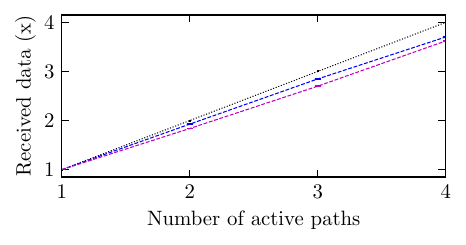}
\caption{Scenario I -- Relative improvement in the amount of received data (blue line -- Click implementation, magenta line -- ns-2 simulation \cite{FAMTAR_CQR}, black dotted line -- linear scaling (for reference)).}
\label{fig:compare}
\end{figure}

Additionally, we compare our results with the simulation results presented in \cite{FAMTAR_CQR}. The comparison is shown in \autoref{fig:compare}. Because the traffic workload used in simulations was different than one in our tests, we compare a relative improvement rather than the absolute amount of received data. The dotted black line shows an ideal linear-scaling. It can be seen, that our FAMTAR implementation achieves even a slightly higher multipath throughput improvement than the simulation, but still slightly below the ideal linear scaling.

% \FloatBarrier

% -----------------------------------------------------
% SCENARIO 2

\subsection*{Scenario II}

This scenario was focused on the verification of FAMTAR under a realistic traffic. The key factor affecting FAMTAR's performance and stability is flow length and size distribution. In the edge case, when all flows in the network are single-packet long, flow caching in FFT becomes meaningless and FAMTAR degrades to a traditional adaptive routing known from ARPANET (which had stability issues).

In order to investigate FAMTAR performance and stability under realistic traffic, we used an UDP flow-based packet generator. The traffic was generated using realistic flow length and size distribution models presented in \cite{flows-agh}. These are the most accurate models currently available in the literature. The flow interarrival rate was set to match 30 Mbit/s of generated traffic in the steady state.

\begin{figure}[h!]
\centering
\includegraphics[width=1.0\columnwidth]{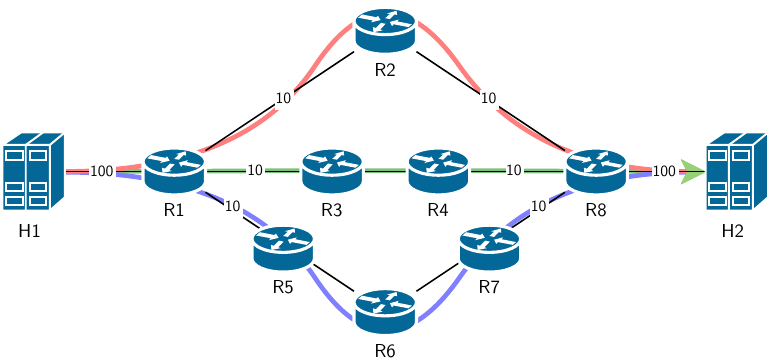}
\caption{Scenario II -- Testing topology.}
\label{fig:scenario-2}
\end{figure}

Moreover, to verify FAMTAR's UCMP capabilities, in this scenario each of three available paths has a different length/cost (\autoref{fig:scenario-2}). In such a scenario, traditional ECMP approaches would be able to use only a single path (the shortest one through the R2). FAMTAR should be able to use all the available paths, starting from the shortest one.

\begin{figure}[h!]
\centering
\includegraphics[scale=0.9]{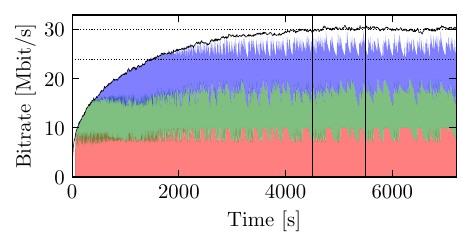}
\caption{Scenario II -- Stackplot of UCMP links utilization. The black solid line shows the generated traffic.}
\label{fig:evaluation-scenario-2-all}
\end{figure}

The \autoref{fig:evaluation-scenario-2-all} contains a stackplot of bitrates of all three paths. Red, green and blue plots correspond to one-, two- and three-hops paths respectively. The black solid line presents the total bitrate of generated traffic.

First of all, it can be seen that it takes more than one hour to reach the steady state by the generator. This is due to the heavy-tailed nature of IP flows. The total throughput of all three paths confirms FAMTAR's UCMP capabilities. FAMTAR can load-balance traffic on the all available paths, even when costs of these paths are different. Moreover, it can be seen that paths with higher cost are started to being used only when all paths with lower costs are congested.

In terms of stability, in the steady state FAMTAR's total throughput varies between 24 and 30 Mbit/s (the dotted horizontal lines on the plot). This is in line with the theory, as 24 Mbit/s corresponds to $ 0.8 $ of all links throughput and $ 0.8 $ is the midpoint value between used congestion thresholds. Simulations presented in previous papers showed a similar results.

\begin{figure}[h!]
\centering
\includegraphics[scale=0.9]{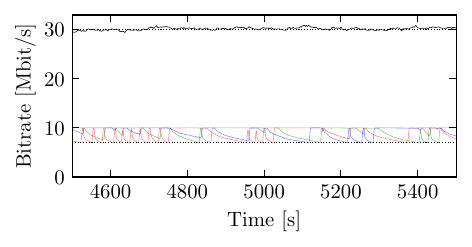}
\caption{Scenario II -- Link loads between 4500 and 5500 second of the experiment, plotted as separate lines. The black solid line shows the generated load.}
\label{fig:evaluation-scenario-2-zoom}
\end{figure}

In \autoref{fig:evaluation-scenario-2-zoom} we present a zoomed-in part of the \autoref{fig:evaluation-scenario-2-all} (between 4500 and 5500 second). In this case plots are not stacked, but presented as separate lines. It can be seen, that traffic oscillations in FAMTAR are significantly reduced thanks to the usage of FFT, which caches routes for existing flows. When congestion occurs, only new flows are routed to the new path, whereas the old ones remain on their existing paths. This introduces an inertia, which reduces both oscillations frequency and their depth. It can be seen that during this 1000-second period, only 25 routing changes happened. The depth of oscillations was limited to $ 0.7 $ of each link throughput, which is the value of $ Th_{min} $.

% \FloatBarrier

% -----------------------------------------------------
% SCENARIO 3

\subsection*{Scenario III}

\begin{table*}[!hb]
  \centering
  \scriptsize
  \caption{Scenario III -- BitTorrent download statistics of the file (Ubuntu ISO CD image, 1 028 653 056 B)} \label{tab:scenario-3}
  \vspace*{6pt}
  \begin{tabular}{@{}lrrr@{}}
    \toprule
      & {Without FAMTAR} & \multicolumn{2}{c}{{With FAMTAR}} \\
      \cmidrule{3-4}
      & & {2 active paths} & {4 active paths} \\
    \midrule
      {Average download time [s]} & $944.3 \pm 8.72$ & $492.7 \pm 15.18$ & $313.0 \pm 22.36$ \\
      {Average download rate [Mbit/s]} & $8.7 \pm 0.08$ & $16.7 \pm 0.52$ & $26.3 \pm 1.88$ \\
      {Relative gain in throughput} & $0\%$ & $92\%$ & $202\%$ \\
    \bottomrule
  \end{tabular}
\end{table*}

\begin{figure}[H]
\centering
\includegraphics[width=1.0\columnwidth]{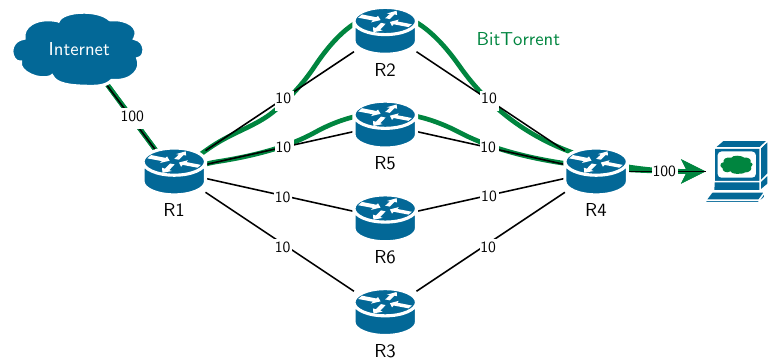}
\caption{Scenario III -- Testing topology.}
\label{fig:scenario-3}
\end{figure}

The goal of the second scenario was to verify the capability of FAMTAR to provide multipath transmission in a~real computer network. A~large file (Ubuntu ISO CD image, 1 GiB) was downloaded through the network using the BitTorrent protocol. During the experiment, router R1 was connected to the Internet, whereas host H2 was downloading the file from external peers (see \autoref{fig:scenario-3}). Connections with peers were established using the TCP and UDP protocols.

We compared a standard IP network to a~FAMTAR-enabled network with $2$ and $4$ parallel paths. The performance was evaluated based on the following metrics:
\begin{itemize}
	\item average download time,
	\item average download rate,
	\item the relative gain in throughput.
\end{itemize}

The results are presented in \autoref{tab:scenario-3}. According to the results, the FAMTAR-enabled network was able to provide much higher transmission speeds (thus shorter download time) than the standard IP network in all considered cases. This means that the FAMTAR-enabled network performed better than the standard IP network with respect to all three considered metrics.

% \FloatBarrier

% -----------------------------------------------------
% SCENARIO 4

\subsection*{Scenario IV}

\begin{figure}[H]
\centering
\includegraphics[width=1.0\columnwidth]{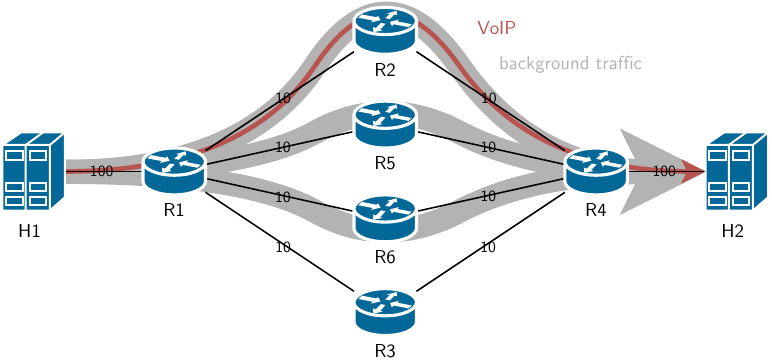}
\caption{Scenario IV -- Testing topology.}
\label{fig:scenario-4}
\end{figure}

The goal of this scenario was to verify the capability of FAMTAR to provide a load-adaptive routing. One important issue related to the presence of congestions in a~network is the effect of background traffic on the Quality of Service (QoS) experienced by certain flows, such as for example related to the Voice-over-IP (VoIP) service. To evaluate the performance of FAMTAR in this context, we started one VoIP flow (G.711.1 RTP) between nodes H1 and H2 (\autoref{fig:scenario-4}), and then the following background flows were scheduled and launched one after another every $200$ ms to gradually consume the available resources on the VoIP flow path:
\begin{itemize}
	\item $50$ UDP flows ($100$ kbit/s, beginning at the $6$th second),
	\item $150$ UDP flows ($100$ kbit/s, beginning at the $25$th second).
\end{itemize}

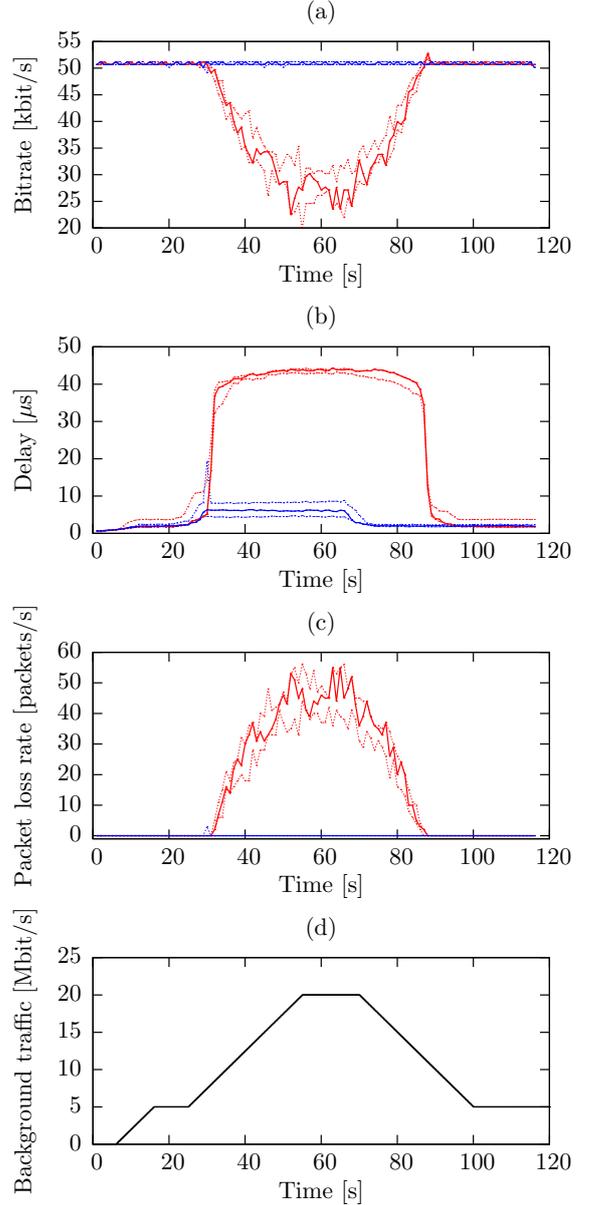
\begin{figure}[H] \center
	\scalebox{0.9}{\import{Plots/scenario-4/}{evaluation-scenario-4}}
	\caption{Scenario IV -- (a) Bitrate, (b) delay, and (c) packet loss rate of the VoIP flow in the congested standard IP and FAMTAR-enabled networks (red and blue solid lines, respectively). The dotted lines reflect the minimum and maximum values acquired in all trials, whereas the solid lines correspond to a single trial. The amount of generated background traffic during the experiment is shown in (d).} \label{fig:evaluation-scenario-4}
\end{figure}

The background flows from the second group were terminated in the same order, with a~$200$ ms interval, starting at the $70$th second. The remaining flows were active until the end of the measurement period. The total amount of generated background traffic during the experiment is presented in \autoref{fig:evaluation-scenario-4}(d).

\autoref{fig:evaluation-scenario-4} show three different QoS-related metrics for the VoIP flow during its activity. We can see that in the case of a standard IP network, packet losses occur in the VoIP flow, starting from the $30$th second when the background traffic begins to exceed $10$ Mbit/s. This observation is in line with our expectations, since the classic IP routing can utilize only one $10$ Mbit/s link. When the offered traffic starts to exceed this level, a portion of packets needs to be dropped in the router's output queue.

In the FAMTAR-enabled network, the VoIP flow remained almost unaffected by the background traffic. The results have proven that FAMTAR eliminates network congestions by redirecting the excessive flows (which would otherwise overload the link) to alternative paths. In the standard IP network, the VoIP stream was mixed with the background traffic forwarded along the same path, and suffered from increased delay, decreased bitrate, and high packet loss rate. We can see it in \autoref{fig:evaluation-scenario-4}(b). Until the $30$th second, the delay in the FAMTAR-enabled network rised in a~similar way as in the standard IP network. This is because the active link was being gradually loaded with the background traffic, so the VoIP packets had to statistically wait longer in the router's queue. However, in the $30$th second, delay stopped increasing in a FAMTAR network. This was caused by the FAMTAR adaptive mechanism which started to use an alternative path for new flows in order to avoid congestion. Consequently, the observed delay remained constant. It started to decrease from the $70$th second when the background flows assigned to the first path started to terminate.

\begin{table}[!h]
  \centering
  \scriptsize
  \caption{Scenario IV -- Effect of traffic congestions on a~VoIP flow} \label{tab:scenario-4}
  \vspace*{6pt}
  \begin{tabular}{@{}lrr@{}}
    \toprule
      & {Without FAMTAR} & {With FAMTAR} \\
    \midrule
      {Min. bitrate [kbit/s]} & $23.04 \pm 2.95$ & $49.97 \pm 0.57$ \\
      {Max. delay [$\mu$s]} & $44 \pm 0.50$ & $9 \pm 7.10$ \\
      {Max. packet loss rate [pkt/s]} & $54.0 \pm 3.62$ & $0.6 \pm 1.67$ \\
    \bottomrule
  \end{tabular}
\end{table}

% \FloatBarrier

% -----------------------------------------------------
% SCENARIO 5

\clearpage

\subsection*{Scenario V}

Transient loops can occur in routing protocols during link cost updates. In traditional approach, such loops are resolved as soon as the state of the routing protocol tables converges on all routers. However, in FAMTAR, routes are being frozen in the FFT when the first packet of a flow passes through the network, which can result in such a transient loop becoming permanent. Such transient loops are resolved immediately by the mechanism presented in \cite{FAMTAR_TTL}.

The same mechanism can also resolve permanent loops, which can occur due to link failures. Such loops are, however, not being resolved immediately. Thus, the goal of the last experiment was to verify the performance and delay of the loop resolution mechanism during a real failure.

\begin{figure}[H]
\centering
\includegraphics[width=1.0\columnwidth]{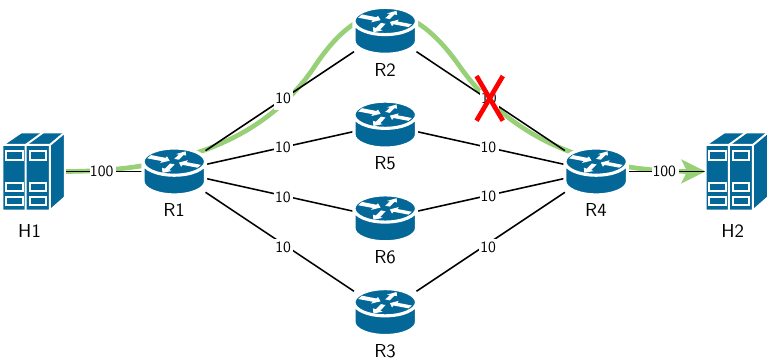}
\caption{Scenario V -- Testing topology}
\label{fig:scenario-5}
\end{figure}

In this scenario, only one flow (constant bitrate of 2.84 Mbit/s, 64 B packets) was transmitted via a~single path between nodes H1 and H2 (\autoref{fig:scenario-5}). This corresponds to approximately 55k packets per second. This means that the amount of traffic did not trigger the multipath mechanism of FAMTAR. We selected a relatively low bitrate to ensure that the results could be compared to the standard IP network.

A failure of the active link was simulated in the network by physically unplugging the network cable from the interface of router R2. It was expected that the network will move the existing flow on a~new path. We compared the number of dropped packets during the restoration phase for both the standard IP and FAMTAR-enabled networks. Traffic generation and measurements were done using the D-ITG software tool. We summarized the results in \autoref{tab:scenario-5}.

\begin{table}[h]
  \centering
  \scriptsize
  \caption{Scenario V -- Number of packets dropped before the traffic was moved on a~new path} \label{tab:scenario-5}
  \vspace*{6pt}
  \begin{tabular}{@{}llll@{}}
    \toprule
      {Without FAMTAR} & {With FAMTAR} & {Avg. difference} & {Relative gain} \\
    \midrule
      $3841.9 \pm 631.5$ & $4656.2 \pm 48.1$ & $-814.3$ & $-21.2 \%$ \\
    \bottomrule
  \end{tabular}
\end{table}

Based on the results, we admit that the deployment of FAMTAR may increase the number of dropped packets during the path restoration phase, which also means that the total restoration time may be longer than in the case of a standard IP network. However, the difference is not large. Having in mind that the generated traffic was 55 kpps, this corresponds to 15 milliseconds of additional outage comparing to classic OSPF routing. Thus, we believe that FAMTAR still demonstrates reasonable performance in the presence of failures. Moreover, the experiment has confirmed that the TTL-based loop resolution mechanism works as expected, i.e., it resolves permanent loops which may appear in the network due to failures.

%%%%%%%%%%%%%%%%%%%%%%%%%%%%%%%%%%%%%%%%%%%%%%%%%%%%%%%%%%%%%%%%%%%%%%%%%%%%%%%%%%%%%%%%%%%%%%%%%%%%%%%%%%%%%%%%%%%%%%%%%%%%%%%%%%%%%%%%%%%%%%%%%%%%%%%%%%%%%%%%%
%%%%%%%%%%%%%%%%%%%%%%%%%%%%%%%%%%%%%%%%%%%%%%%%%%%%%%%%%%%%%%%%%%%%%%%%%%%%%%%%%%%%%%%%%%%%%%%%%%%%%%%%%%%%%%%%%%%%%%%%%%%%%%%%%%%%%%%%%%%%%%%%%%%%%%%%%%%%%%%%%
%%%%%%%%%%%%%%%%%%%%%%%%%%%%%%%%%%%%%%%%%%%%%%%%%%%%%%%%%%%%%%%%%%%%%%%%%%%%%%%%%%%%%%%%%%%%%%%%%%%%%%%%%%%%%%%%%%%%%%%%%%%%%%%%%%%%%%%%%%%%%%%%%%%%%%%%%%%%%%%%%
\section{Conclusion} \label{sec:conclusion}
FAMTAR is a promising concept for multipath routing in IP networks. So far, the approach was presented and evaluated through simulations. Now, we have a prototype which we tested in physical networks.

We presented the practical aspects of a prototype implementation. We also documented and solved challenges and problems encountered during the implementation process, which were not noticed during simulations. After basic tests verifying the proper operation of each component, we benchmarked physical networks built with the FAMTAR routers under real traffic conditions. The results show significant advantages of FAMTAR compared to standard IP routing. Most importantly, the test results are in line with simulation analysis published earlier.

We believe that the implementation and test results presented in this paper will accelerate further development of FAMTAR. Now, FAMTAR is a robust solution which is also easy to implement.

%%%%%%%%%%%%%%%%%%%%%%%%%%%%%%%%%%%%%%%%%%%%%%%%%%%%%%%%%%%%%%%%%%%%%%%%%%%%%%%%%%%%%%%%%%%%%%%%%%%%%%%%%%%%%%%%%%%%%%%%%%%%%%%%%%%%%%%%%%%%%%%%%%%%%%%%%%%%%%%%%
%%%%%%%%%%%%%%%%%%%%%%%%%%%%%%%%%%%%%%%%%%%%%%%%%%%%%%%%%%%%%%%%%%%%%%%%%%%%%%%%%%%%%%%%%%%%%%%%%%%%%%%%%%%%%%%%%%%%%%%%%%%%%%%%%%%%%%%%%%%%%%%%%%%%%%%%%%%%%%%%%
%%%%%%%%%%%%%%%%%%%%%%%%%%%%%%%%%%%%%%%%%%%%%%%%%%%%%%%%%%%%%%%%%%%%%%%%%%%%%%%%%%%%%%%%%%%%%%%%%%%%%%%%%%%%%%%%%%%%%%%%%%%%%%%%%%%%%%%%%%%%%%%%%%%%%%%%%%%%%%%%%
\section*{Acknowledgment}
\noindent
The research was carried out with the support of the project "Flow-Aware Multi-Topology Adaptive Routing" funded by the National Centre for Research and Development in Poland under the LIDER programme.

%%%%%%%%%%%%%%%%%%%%%%%%%%%%%%%%%%%%%%%%%%%%%%%%%%%%%%%%%%%%%%%%%%%%%%%%%%%%%%%%%%%%%%%%%%%%%%%%%%%%%%%%%%%%%%%%%%%%%%%%%%%%%%%%%%%%%%%%%%%%%%%%%%%%%%%%%%%%%%%%%
%%%%%%%%%%%%%%%%%%%%%%%%%%%%%%%%%%%%%%%%%%%%%%%%%%%%%%%%%%%%%%%%%%%%%%%%%%%%%%%%%%%%%%%%%%%%%%%%%%%%%%%%%%%%%%%%%%%%%%%%%%%%%%%%%%%%%%%%%%%%%%%%%%%%%%%%%%%%%%%%%
%%%%%%%%%%%%%%%%%%%%%%%%%%%%%%%%%%%%%%%%%%%%%%%%%%%%%%%%%%%%%%%%%%%%%%%%%%%%%%%%%%%%%%%%%%%%%%%%%%%%%%%%%%%%%%%%%%%%%%%%%%%%%%%%%%%%%%%%%%%%%%%%%%%%%%%%%%%%%%%%%

\bibliography{./Bibliography/mybib}
\bibliographystyle{elsarticle-num}

\end{document}

%% file: Plots/scenario-1/evaluation-scenario-1.tex
% GNUPLOT: LaTeX picture with Postscript
\begingroup
  \makeatletter
  \providecommand\color[2][]{%
    \GenericError{(gnuplot) \space\space\space\@spaces}{%
      Package color not loaded in conjunction with
      terminal option `colourtext'%
    }{See the gnuplot documentation for explanation.%
    }{Either use 'blacktext' in gnuplot or load the package
      color.sty in LaTeX.}%
    \renewcommand\color[2][]{}%
  }%
  \providecommand\includegraphics[2][]{%
    \GenericError{(gnuplot) \space\space\space\@spaces}{%
      Package graphicx or graphics not loaded%
    }{See the gnuplot documentation for explanation.%
    }{The gnuplot epslatex terminal needs graphicx.sty or graphics.sty.}%
    \renewcommand\includegraphics[2][]{}%
  }%
  \providecommand\rotatebox[2]{#2}%
  \@ifundefined{ifGPcolor}{%
    \newif\ifGPcolor
    \GPcolortrue
  }{}%
  \@ifundefined{ifGPblacktext}{%
    \newif\ifGPblacktext
    \GPblacktexttrue
  }{}%
  % define a \g@addto@macro without @ in the name:
  \let\gplgaddtomacro\g@addto@macro
  % define empty templates for all commands taking text:
  \gdef\gplbacktext{}%
  \gdef\gplfronttext{}%
  \makeatother
  \ifGPblacktext
    % no textcolor at all
    \def\colorrgb#1{}%
    \def\colorgray#1{}%
  \else
    % gray or color?
    \ifGPcolor
      \def\colorrgb#1{\color[rgb]{#1}}%
      \def\colorgray#1{\color[gray]{#1}}%
      \expandafter\def\csname LTw\endcsname{\color{white}}%
      \expandafter\def\csname LTb\endcsname{\color{black}}%
      \expandafter\def\csname LTa\endcsname{\color{black}}%
      \expandafter\def\csname LT0\endcsname{\color[rgb]{1,0,0}}%
      \expandafter\def\csname LT1\endcsname{\color[rgb]{0,1,0}}%
      \expandafter\def\csname LT2\endcsname{\color[rgb]{0,0,1}}%
      \expandafter\def\csname LT3\endcsname{\color[rgb]{1,0,1}}%
      \expandafter\def\csname LT4\endcsname{\color[rgb]{0,1,1}}%
      \expandafter\def\csname LT5\endcsname{\color[rgb]{1,1,0}}%
      \expandafter\def\csname LT6\endcsname{\color[rgb]{0,0,0}}%
      \expandafter\def\csname LT7\endcsname{\color[rgb]{1,0.3,0}}%
      \expandafter\def\csname LT8\endcsname{\color[rgb]{0.5,0.5,0.5}}%
    \else
      % gray
      \def\colorrgb#1{\color{black}}%
      \def\colorgray#1{\color[gray]{#1}}%
      \expandafter\def\csname LTw\endcsname{\color{white}}%
      \expandafter\def\csname LTb\endcsname{\color{black}}%
      \expandafter\def\csname LTa\endcsname{\color{black}}%
      \expandafter\def\csname LT0\endcsname{\color{black}}%
      \expandafter\def\csname LT1\endcsname{\color{black}}%
      \expandafter\def\csname LT2\endcsname{\color{black}}%
      \expandafter\def\csname LT3\endcsname{\color{black}}%
      \expandafter\def\csname LT4\endcsname{\color{black}}%
      \expandafter\def\csname LT5\endcsname{\color{black}}%
      \expandafter\def\csname LT6\endcsname{\color{black}}%
      \expandafter\def\csname LT7\endcsname{\color{black}}%
      \expandafter\def\csname LT8\endcsname{\color{black}}%
    \fi
  \fi
  \setlength{\unitlength}{0.0500bp}%
  \begin{picture}(4761.40,12755.00)%
    \gplgaddtomacro\gplbacktext{%
      \csname LTb\endcsname%
      \put(688,10716){\makebox(0,0)[r]{\strut{} 200}}%
      \put(688,10938){\makebox(0,0)[r]{\strut{} 300}}%
      \put(688,11161){\makebox(0,0)[r]{\strut{} 400}}%
      \put(688,11383){\makebox(0,0)[r]{\strut{} 500}}%
      \put(688,11606){\makebox(0,0)[r]{\strut{} 600}}%
      \put(688,11828){\makebox(0,0)[r]{\strut{} 700}}%
      \put(688,12051){\makebox(0,0)[r]{\strut{} 800}}%
      \put(688,12273){\makebox(0,0)[r]{\strut{} 900}}%
      \put(784,10556){\makebox(0,0){\strut{} 1}}%
      \put(2013,10556){\makebox(0,0){\strut{} 2}}%
      \put(3243,10556){\makebox(0,0){\strut{} 3}}%
      \put(4472,10556){\makebox(0,0){\strut{} 4}}%
      \put(128,11494){\rotatebox{-270}{\makebox(0,0){\strut{}Received data [MB]}}}%
      \put(2628,10316){\makebox(0,0){\strut{}Number of active paths}}%
      \put(2628,12513){\makebox(0,0){\strut{}(a)}}%
    }%
    \gplgaddtomacro\gplfronttext{%
    }%
    \gplgaddtomacro\gplbacktext{%
      \csname LTb\endcsname%
      \put(592,8165){\makebox(0,0)[r]{\strut{} 10}}%
      \put(592,8387){\makebox(0,0)[r]{\strut{} 20}}%
      \put(592,8610){\makebox(0,0)[r]{\strut{} 30}}%
      \put(592,8832){\makebox(0,0)[r]{\strut{} 40}}%
      \put(592,9055){\makebox(0,0)[r]{\strut{} 50}}%
      \put(592,9277){\makebox(0,0)[r]{\strut{} 60}}%
      \put(592,9500){\makebox(0,0)[r]{\strut{} 70}}%
      \put(592,9722){\makebox(0,0)[r]{\strut{} 80}}%
      \put(688,8005){\makebox(0,0){\strut{} 1}}%
      \put(1949,8005){\makebox(0,0){\strut{} 2}}%
      \put(3211,8005){\makebox(0,0){\strut{} 3}}%
      \put(4472,8005){\makebox(0,0){\strut{} 4}}%
      \put(128,8943){\rotatebox{-270}{\makebox(0,0){\strut{}Packets dropped ($\%$)}}}%
      \put(2580,7765){\makebox(0,0){\strut{}Number of active paths}}%
      \put(2580,9962){\makebox(0,0){\strut{}(b)}}%
    }%
    \gplgaddtomacro\gplfronttext{%
    }%
    \gplgaddtomacro\gplbacktext{%
      \csname LTb\endcsname%
      \put(592,5744){\makebox(0,0)[r]{\strut{} 28}}%
      \put(592,6004){\makebox(0,0)[r]{\strut{} 32}}%
      \put(592,6263){\makebox(0,0)[r]{\strut{} 36}}%
      \put(592,6523){\makebox(0,0)[r]{\strut{} 40}}%
      \put(592,6783){\makebox(0,0)[r]{\strut{} 44}}%
      \put(592,7042){\makebox(0,0)[r]{\strut{} 48}}%
      \put(688,5454){\makebox(0,0){\strut{} 1}}%
      \put(1949,5454){\makebox(0,0){\strut{} 2}}%
      \put(3211,5454){\makebox(0,0){\strut{} 3}}%
      \put(4472,5454){\makebox(0,0){\strut{} 4}}%
      \put(128,6393){\rotatebox{-270}{\makebox(0,0){\strut{}Average delay [ms]}}}%
      \put(2580,5214){\makebox(0,0){\strut{}Number of active paths}}%
      \put(2580,7412){\makebox(0,0){\strut{}(c)}}%
    }%
    \gplgaddtomacro\gplfronttext{%
    }%
    \gplgaddtomacro\gplbacktext{%
      \csname LTb\endcsname%
      \put(592,3063){\makebox(0,0)[r]{\strut{} 0}}%
      \put(592,3375){\makebox(0,0)[r]{\strut{} 10}}%
      \put(592,3686){\makebox(0,0)[r]{\strut{} 20}}%
      \put(592,3998){\makebox(0,0)[r]{\strut{} 30}}%
      \put(592,4309){\makebox(0,0)[r]{\strut{} 40}}%
      \put(592,4621){\makebox(0,0)[r]{\strut{} 50}}%
      \put(688,2903){\makebox(0,0){\strut{} 1}}%
      \put(1949,2903){\makebox(0,0){\strut{} 2}}%
      \put(3211,2903){\makebox(0,0){\strut{} 3}}%
      \put(4472,2903){\makebox(0,0){\strut{} 4}}%
      \put(128,3842){\rotatebox{-270}{\makebox(0,0){\strut{}Minimum delay [ms]}}}%
      \put(2580,2663){\makebox(0,0){\strut{}Number of active paths}}%
      \put(2580,4861){\makebox(0,0){\strut{}(d)}}%
    }%
    \gplgaddtomacro\gplfronttext{%
    }%
    \gplgaddtomacro\gplbacktext{%
      \csname LTb\endcsname%
      \put(688,512){\makebox(0,0)[r]{\strut{} 0}}%
      \put(688,707){\makebox(0,0)[r]{\strut{} 100}}%
      \put(688,902){\makebox(0,0)[r]{\strut{} 200}}%
      \put(688,1097){\makebox(0,0)[r]{\strut{} 300}}%
      \put(688,1292){\makebox(0,0)[r]{\strut{} 400}}%
      \put(688,1486){\makebox(0,0)[r]{\strut{} 500}}%
      \put(688,1681){\makebox(0,0)[r]{\strut{} 600}}%
      \put(688,1876){\makebox(0,0)[r]{\strut{} 700}}%
      \put(688,2071){\makebox(0,0)[r]{\strut{} 800}}%
      \put(784,352){\makebox(0,0){\strut{} 1}}%
      \put(2013,352){\makebox(0,0){\strut{} 2}}%
      \put(3243,352){\makebox(0,0){\strut{} 3}}%
      \put(4472,352){\makebox(0,0){\strut{} 4}}%
      \put(128,1291){\rotatebox{-270}{\makebox(0,0){\strut{}Maximum delay [ms]}}}%
      \put(2628,112){\makebox(0,0){\strut{}Number of active paths}}%
      \put(2628,2311){\makebox(0,0){\strut{}(e)}}%
    }%
    \gplgaddtomacro\gplfronttext{%
    }%
    \gplbacktext
    \put(0,0){\includegraphics{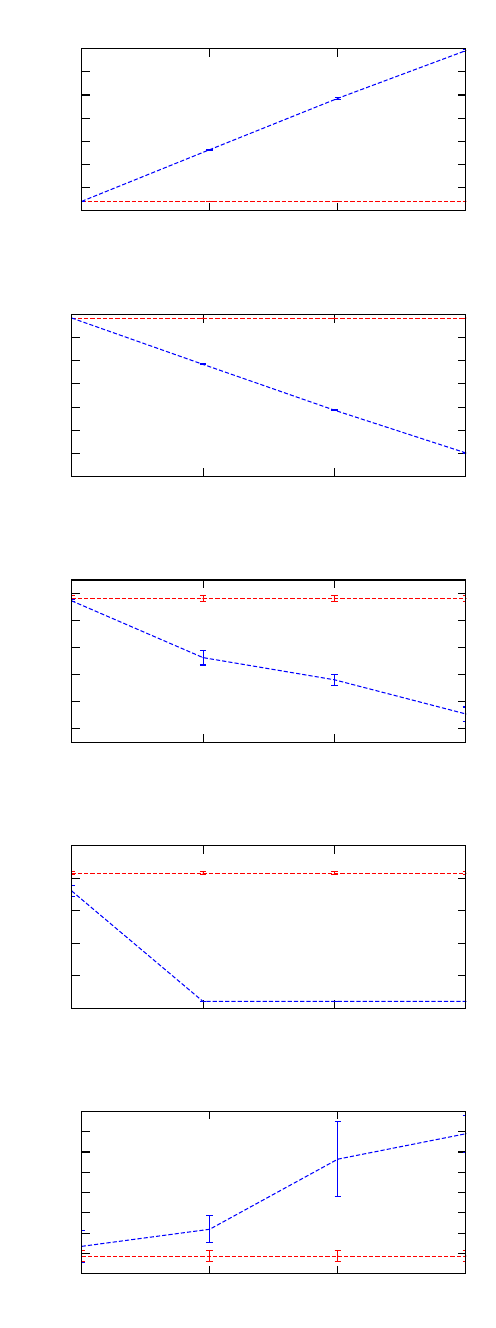}}%
    \gplfronttext
  \end{picture}%
\endgroup

%% file: Plots/scenario-4/evaluation-scenario-4.tex
% GNUPLOT: LaTeX picture with Postscript
\begingroup
  \makeatletter
  \providecommand\color[2][]{%
    \GenericError{(gnuplot) \space\space\space\@spaces}{%
      Package color not loaded in conjunction with
      terminal option `colourtext'%
    }{See the gnuplot documentation for explanation.%
    }{Either use 'blacktext' in gnuplot or load the package
      color.sty in LaTeX.}%
    \renewcommand\color[2][]{}%
  }%
  \providecommand\includegraphics[2][]{%
    \GenericError{(gnuplot) \space\space\space\@spaces}{%
      Package graphicx or graphics not loaded%
    }{See the gnuplot documentation for explanation.%
    }{The gnuplot epslatex terminal needs graphicx.sty or graphics.sty.}%
    \renewcommand\includegraphics[2][]{}%
  }%
  \providecommand\rotatebox[2]{#2}%
  \@ifundefined{ifGPcolor}{%
    \newif\ifGPcolor
    \GPcolortrue
  }{}%
  \@ifundefined{ifGPblacktext}{%
    \newif\ifGPblacktext
    \GPblacktexttrue
  }{}%
  % define a \g@addto@macro without @ in the name:
  \let\gplgaddtomacro\g@addto@macro
  % define empty templates for all commands taking text:
  \gdef\gplbacktext{}%
  \gdef\gplfronttext{}%
  \makeatother
  \ifGPblacktext
    % no textcolor at all
    \def\colorrgb#1{}%
    \def\colorgray#1{}%
  \else
    % gray or color?
    \ifGPcolor
      \def\colorrgb#1{\color[rgb]{#1}}%
      \def\colorgray#1{\color[gray]{#1}}%
      \expandafter\def\csname LTw\endcsname{\color{white}}%
      \expandafter\def\csname LTb\endcsname{\color{black}}%
      \expandafter\def\csname LTa\endcsname{\color{black}}%
      \expandafter\def\csname LT0\endcsname{\color[rgb]{1,0,0}}%
      \expandafter\def\csname LT1\endcsname{\color[rgb]{0,1,0}}%
      \expandafter\def\csname LT2\endcsname{\color[rgb]{0,0,1}}%
      \expandafter\def\csname LT3\endcsname{\color[rgb]{1,0,1}}%
      \expandafter\def\csname LT4\endcsname{\color[rgb]{0,1,1}}%
      \expandafter\def\csname LT5\endcsname{\color[rgb]{1,1,0}}%
      \expandafter\def\csname LT6\endcsname{\color[rgb]{0,0,0}}%
      \expandafter\def\csname LT7\endcsname{\color[rgb]{1,0.3,0}}%
      \expandafter\def\csname LT8\endcsname{\color[rgb]{0.5,0.5,0.5}}%
    \else
      % gray
      \def\colorrgb#1{\color{black}}%
      \def\colorgray#1{\color[gray]{#1}}%
      \expandafter\def\csname LTw\endcsname{\color{white}}%
      \expandafter\def\csname LTb\endcsname{\color{black}}%
      \expandafter\def\csname LTa\endcsname{\color{black}}%
      \expandafter\def\csname LT0\endcsname{\color{black}}%
      \expandafter\def\csname LT1\endcsname{\color{black}}%
      \expandafter\def\csname LT2\endcsname{\color{black}}%
      \expandafter\def\csname LT3\endcsname{\color{black}}%
      \expandafter\def\csname LT4\endcsname{\color{black}}%
      \expandafter\def\csname LT5\endcsname{\color{black}}%
      \expandafter\def\csname LT6\endcsname{\color{black}}%
      \expandafter\def\csname LT7\endcsname{\color{black}}%
      \expandafter\def\csname LT8\endcsname{\color{black}}%
    \fi
  \fi
  \setlength{\unitlength}{0.0500bp}%
  \begin{picture}(4761.40,10204.00)%
    \gplgaddtomacro\gplbacktext{%
      \csname LTb\endcsname%
      \put(592,8165){\makebox(0,0)[r]{\strut{} 20}}%
      \put(592,8387){\makebox(0,0)[r]{\strut{} 25}}%
      \put(592,8610){\makebox(0,0)[r]{\strut{} 30}}%
      \put(592,8832){\makebox(0,0)[r]{\strut{} 35}}%
      \put(592,9055){\makebox(0,0)[r]{\strut{} 40}}%
      \put(592,9277){\makebox(0,0)[r]{\strut{} 45}}%
      \put(592,9500){\makebox(0,0)[r]{\strut{} 50}}%
      \put(592,9722){\makebox(0,0)[r]{\strut{} 55}}%
      \put(688,8005){\makebox(0,0){\strut{} 0}}%
      \put(1319,8005){\makebox(0,0){\strut{} 20}}%
      \put(1949,8005){\makebox(0,0){\strut{} 40}}%
      \put(2580,8005){\makebox(0,0){\strut{} 60}}%
      \put(3211,8005){\makebox(0,0){\strut{} 80}}%
      \put(3841,8005){\makebox(0,0){\strut{} 100}}%
      \put(4472,8005){\makebox(0,0){\strut{} 120}}%
      \put(128,8943){\rotatebox{-270}{\makebox(0,0){\strut{}Bitrate [kbit/s]}}}%
      \put(2580,7765){\makebox(0,0){\strut{}Time [s]}}%
      \put(2580,9962){\makebox(0,0){\strut{}(a)}}%
    }%
    \gplgaddtomacro\gplfronttext{%
    }%
    \gplgaddtomacro\gplbacktext{%
      \csname LTb\endcsname%
      \put(592,5614){\makebox(0,0)[r]{\strut{} 0}}%
      \put(592,5926){\makebox(0,0)[r]{\strut{} 10}}%
      \put(592,6237){\makebox(0,0)[r]{\strut{} 20}}%
      \put(592,6549){\makebox(0,0)[r]{\strut{} 30}}%
      \put(592,6860){\makebox(0,0)[r]{\strut{} 40}}%
      \put(592,7172){\makebox(0,0)[r]{\strut{} 50}}%
      \put(688,5454){\makebox(0,0){\strut{} 0}}%
      \put(1319,5454){\makebox(0,0){\strut{} 20}}%
      \put(1949,5454){\makebox(0,0){\strut{} 40}}%
      \put(2580,5454){\makebox(0,0){\strut{} 60}}%
      \put(3211,5454){\makebox(0,0){\strut{} 80}}%
      \put(3841,5454){\makebox(0,0){\strut{} 100}}%
      \put(4472,5454){\makebox(0,0){\strut{} 120}}%
      \put(128,6393){\rotatebox{-270}{\makebox(0,0){\strut{}Delay [$\mu$s]}}}%
      \put(2580,5214){\makebox(0,0){\strut{}Time [s]}}%
      \put(2580,7412){\makebox(0,0){\strut{}(b)}}%
    }%
    \gplgaddtomacro\gplfronttext{%
    }%
    \gplgaddtomacro\gplbacktext{%
      \csname LTb\endcsname%
      \put(592,3089){\makebox(0,0)[r]{\strut{} 0}}%
      \put(592,3344){\makebox(0,0)[r]{\strut{} 10}}%
      \put(592,3599){\makebox(0,0)[r]{\strut{} 20}}%
      \put(592,3855){\makebox(0,0)[r]{\strut{} 30}}%
      \put(592,4110){\makebox(0,0)[r]{\strut{} 40}}%
      \put(592,4366){\makebox(0,0)[r]{\strut{} 50}}%
      \put(592,4621){\makebox(0,0)[r]{\strut{} 60}}%
      \put(688,2903){\makebox(0,0){\strut{} 0}}%
      \put(1319,2903){\makebox(0,0){\strut{} 20}}%
      \put(1949,2903){\makebox(0,0){\strut{} 40}}%
      \put(2580,2903){\makebox(0,0){\strut{} 60}}%
      \put(3211,2903){\makebox(0,0){\strut{} 80}}%
      \put(3841,2903){\makebox(0,0){\strut{} 100}}%
      \put(4472,2903){\makebox(0,0){\strut{} 120}}%
      \put(128,3842){\rotatebox{-270}{\makebox(0,0){\strut{}Packet loss rate [packets/s]}}}%
      \put(2580,2663){\makebox(0,0){\strut{}Time [s]}}%
      \put(2580,4861){\makebox(0,0){\strut{}(c)}}%
    }%
    \gplgaddtomacro\gplfronttext{%
    }%
    \gplgaddtomacro\gplbacktext{%
      \csname LTb\endcsname%
      \put(592,512){\makebox(0,0)[r]{\strut{} 0}}%
      \put(592,824){\makebox(0,0)[r]{\strut{} 5}}%
      \put(592,1136){\makebox(0,0)[r]{\strut{} 10}}%
      \put(592,1447){\makebox(0,0)[r]{\strut{} 15}}%
      \put(592,1759){\makebox(0,0)[r]{\strut{} 20}}%
      \put(592,2071){\makebox(0,0)[r]{\strut{} 25}}%
      \put(688,352){\makebox(0,0){\strut{} 0}}%
      \put(1319,352){\makebox(0,0){\strut{} 20}}%
      \put(1949,352){\makebox(0,0){\strut{} 40}}%
      \put(2580,352){\makebox(0,0){\strut{} 60}}%
      \put(3211,352){\makebox(0,0){\strut{} 80}}%
      \put(3841,352){\makebox(0,0){\strut{} 100}}%
      \put(4472,352){\makebox(0,0){\strut{} 120}}%
      \put(128,1291){\rotatebox{-270}{\makebox(0,0){\strut{}Background traffic [Mbit/s]}}}%
      \put(2580,112){\makebox(0,0){\strut{}Time [s]}}%
      \put(2580,2311){\makebox(0,0){\strut{}(d)}}%
    }%
    \gplgaddtomacro\gplfronttext{%
    }%
    \gplbacktext
    \put(0,0){\includegraphics{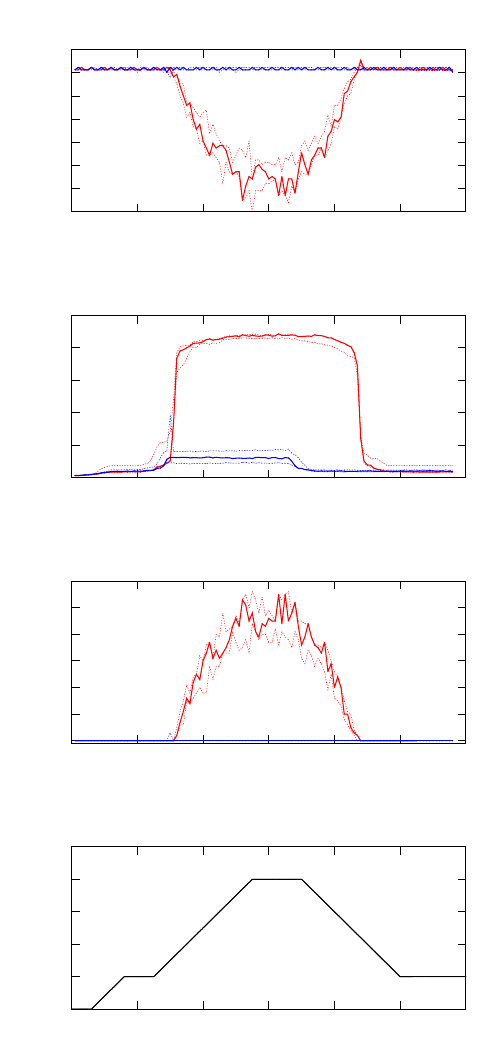}}%
    \gplfronttext
  \end{picture}%
\endgroup